\documentclass[debug]{rmaa}

\usepackage{paralist}

\usepackage{psfrag,color}
\usepackage[latin1]{inputenc}
\usepackage{caption}
\usepackage{subcaption}
\usepackage{natbib}
\usepackage{multirow}
\usepackage[flushleft]{threeparttable}
\usepackage{booktabs,caption}

\title{A DEEP STUDY OF OPEN CLUSTER NGC 5288
USING PHOTOMETRIC AND ASTROMETRIC DATA FROM GAIA DR3 AND 2MASS.} 

\author{Ritika Sethi\altaffilmark{1}, D. Bisht$^{*}$\altaffilmark{2}, Geeta Rangwal\altaffilmark{3}, A. Raj\altaffilmark{2}
  \altaffilmark{}}

\altaffiltext{1}{Department of Physical Sciences, Indian Institute of Science Education and Research, Berhampur, Odisha 760010, India }
\altaffiltext{2}{Indian Centre for Space Physics, 466 Barakhola, Singabari road, Netai Nagar, Kolkata 700099, India}
\altaffiltext{3}{Indian Institute of Astrophysics, Koramangala II BLock, Bangalore, 560034, India}

\shortauthor{Ritika Sethi et al.}
\shorttitle{Deep Study of NGC 5288 using Gaia DR3 data.}

\fulladdresses{}

\listofauthors{Ritika Sethi, D. Bisht, Geeta Rangawal}
\indexauthor{Sethi, Ritika}
\indexauthor{Collaborator, A.}

\abstract{This paper investigates a poorly studied open cluster, NGC 5288, using 2MASS \textit{JHK$_{S}$} and the recently released Gaia DR3 astrometric and photometric data. The mean proper motions in Right Ascension and Declination are estimated as $(-3.840 \pm 0.230)$ and $(-1.934 \pm 0.162)$ mas yr$^{-1}$ respectively. We also derive the age and distance of the cluster as $510 \pm 190$ Myr and $2.64 \pm 0.11$ kpc,  using color-magnitude diagrams (CMDs). We have also obtained distance as $2.77\pm0.42$ kpc using the parallax method. Interstellar reddening E(B-V) in the direction of the cluster is determined as $0.45$ mag using the ($(J-H), (J-K)$) color-color diagram. We have found the mass function slope for main-sequence stars as $1.39\pm0.29$ within the mass range 1.0$-$2.7 $M_\odot$, which agrees with Salpeter's value within uncertainty. Galactic orbits are derived using the Galactic potential model, indicating that NGC 5288 follows a circular path around the Galactic center.}

\resumen{}

\addkeyword{Colour Magnitude Diagrams}
\addkeyword{Initial Mass Function}
\addkeyword{Luminosity Function}
\addkeyword{Membership Probability}
\addkeyword{Open Cluste}
\addkeyword{Proper motion}

\begin{document}
\maketitle

\section{Introduction}
\label{sec:intro}
Open clusters (OCs) are loosely bound systems consisting of around a hundred to a few thousand stars. Owing to the origin of cluster members from a common molecular cloud, they have similar chemical composition, kinematic properties and distance from the Sun \citep{2002yCat.7229....0D, 2016yCat..35930116J}. This makes them prime candidates for studying stellar formation and evolution. OCs can help us conveniently determine various astrophysical parameters and provide insights into the chemical and structural evolution of the Milky Way Galaxy (Sariya et al. 2021a; Tadross \& Hendy 2021).

 Global space missions like Gaia, launched in 2013, have provided the scientific community with high-precision, 3-dimensional spatial and velocity distributions of more than 3 billion stars \citep{2016A&A...595A...1G}. The mission has transformed the field of astrophysics and established a foundation for high-quality research. Although it is challenging to distinguish open cluster members from the background-field stars, the 3-D data from the Gaia mission has made the job easier \citep{2022gdr3.reptE....V}. Much work has recently been done to estimate the membership probability of stars in the cluster region using the proper motion and parallax information (Cantat-Gaudin et al. 2018; Bisht et al. 2019; Ferreira et al. 2020; Sariya et al. 2021a). The decontaminated sample of cluster members thus obtained, naturally yield cleaner color-magnitude diagrams and more accurate estimates of the astrophysical parameters. 

In this study, we estimate the astrophysical properties of NGC 5288, an open cluster positioned at $\alpha_{2000} = 13^h 48^m 44^s$, $\delta_{2000} = -64^{\circ}41'06''$ corresponding to the Galactic coordinates $l = 309^\circ.010$ and $b = -2^\circ.492$ according to the WEBDA open cluster database. It is located in the constellation Circinus and situated to the south of the celestial equator, mostly visible from the southern hemisphere. It is designated as Cr 278 in the Collinder catalogue of open clusters \citep{1931AnLun...2....1C}. The charge-coupled device (CCD) photometric data of the cluster observed at the Cerro Tololo Inter-American Observatory (CTIO) was analysed by \citet{2006MNRAS.367..599P} for determining fundamental parameters like size, age, reddening, distance and metallicity. They described NGC 5288 as a rich, strongly absorbed cluster with a small but bright nucleus and a low-density extended corona located beyond the Carina spiral feature. However, the extensive scattering caused due to atmospheric disturbances and low resolution of photometric data from ground based telescopes make it difficult to study individual stars in the cluster, differentiate them against the background field which restricts scientists to carry out further research. The space based astrometric solutions from Gaia mission are hence of immense importance for modern research with more accuracy. Specifically for NGC 5288, the Gaia data helped us estimate the basic parameters of the cluster more accurately. Also completeness of Gaia photometric data is 100 percent for this cluster for a limit of 20 mag, which further helped us in finding the actual present day Mass function slope of the cluster.

This paper is organized as follows. In Section \ref{2}, we describe the data used, and in Section \ref{3}, we elaborate on the procedure to calculate the membership probability of the stars. In Section \ref{4}, we derive the cluster's structural properties and basic parameters. Section \ref{5} sheds light on fitting the derived color magnitude diagram (CMD) of the cluster on theoretical isochrones and subsequently evaluating its age and distance. Section \ref{6} discusses the luminosity and mass functions of the cluster in detail. Section \ref{7} is devoted to the orbital dynamics of the cluster. Finally, in Section \ref{8}, we conclude the study and present a summary of the results.

\section{Data Used} \label{2}

\subsection{2MASS}

In this paper, we have used The two micros all-sky ($2MASS$, Skrutskie et al. 2006) survey data for the cluster NGC 5288. This data sets use two highly automated 1.3m telescopes, one at Mt. Hopkins, Arizona (AZ),
USA and the other at CTIO, Chile, with a 3-channel camera (256$\times$256 array of HgCdTe detectors in each channel).
$2MASS$ catalog provides $J~ (1.25~ \mu m)$, $H~ (1.65~ \mu m)$ and $K_{s}~ (2.17~ \mu m)$ band photometry for millions
of galaxies and nearly a half-billion stars (Carpenter, 2001). The sensitivity of this catalog is 15.8 mag for $J$,
15.1 mag for $H$ and 14.3 mag for $K_{s}$ band at $S/N$=10.

\subsection{GAIA DR3}
The Gaia DR3 data gives a complete astrometric solution for more than $1.46$ billion sources. It consists of their positions on the sky ($\alpha, \delta$), parallaxes, and proper motions within 3 to 21 mag limit in the G band \citep{2016A&A...595A...1G, 2022arXiv220605989B}. The Gaia photometric bands: G, G$_{BP}$, and G$_{RP}$ gather measurements over the wavelength ranges of $330-1050, 330-680$, and $640-1050$ nm, respectively.

\begin{figure}[!t]
  \includegraphics[width=\columnwidth]{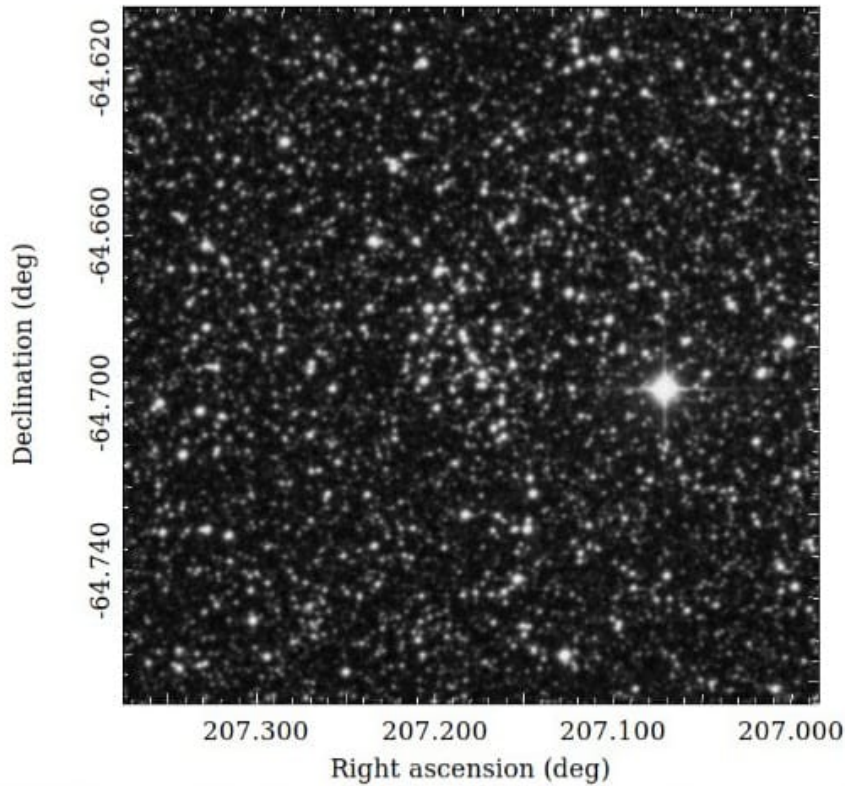}
  \caption{Identification map of NGC 5288 taken from DSS. }
  \label{fig1}
\end{figure}

\begin{figure}[!t]
  \includegraphics[width=\columnwidth]{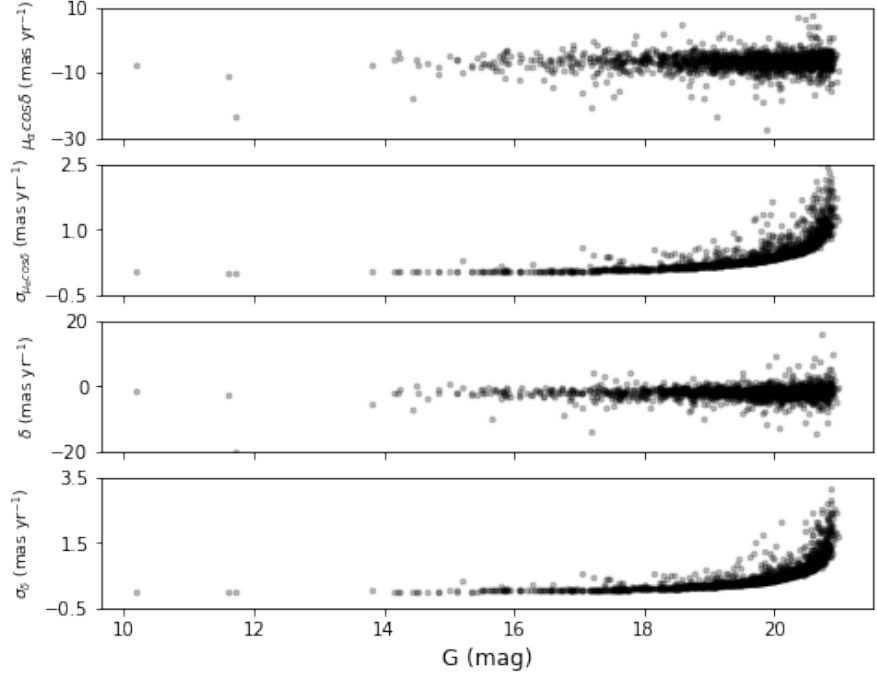}
  \caption{Plot of proper motions and their errors against G mag}
  \label{fig2}
\end{figure}

We collected the data within a radius of 10 arcmins, taking into account the estimated cluster radius of 6.3 arcmins from an earlier study \citep{2006MNRAS.367..599P} of the cluster. The identification map of the cluster under study is shown in Fig. \ref{fig1} and is taken from Digitized Sky Survey (DSS). The proper motions and their respective errors of all stars in the Gaia data base are plotted against G mag in Fig \ref{fig2}. 

\section{Identification of Cluster Members}
\label{3}
\subsection{Vector Point Diagrams}
\begin{figure}[!t]
  \includegraphics[width=\columnwidth]{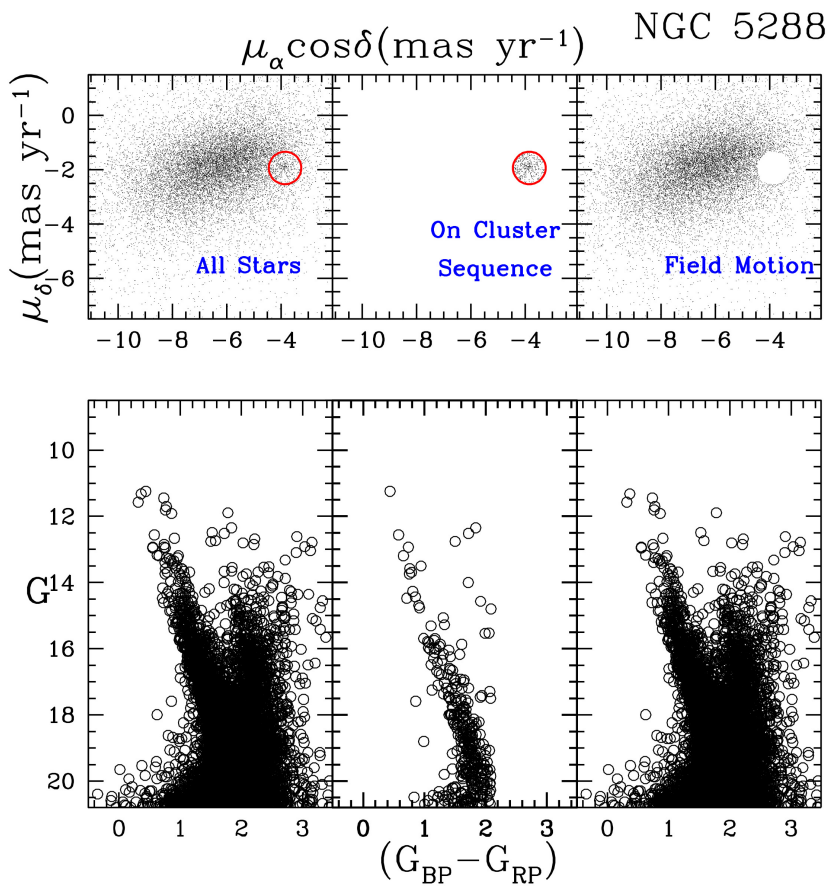}
  \caption{(Top panel) VPDs for cluster NGC 5288. (Bottom panel) $G$ vs ($G_{BP} - G_{RP}$) CMDs of the stars shown in the VPD above them. (Left panel) includes all stars. (Middle panel) includes only the stars within the region inside the red circle; which we classify as the probable cluster members. (Right panel) includes the probable background/foreground field stars).}
  \label{fig3}
\end{figure}
As mentioned above, the cluster stars share similar kinematic properties and the same mean motions. Therefore, proper motions of stars are a crucial parameter for identifying cluster members in field stars (Yadav et al. 2013; Bisht et al. 2021b). A graph is plotted between the proper motion components of the stars, $\mu_\alpha \cos\delta$ mas yr$^{-1}$ and $\mu_\delta$ mas yr$^{-1}$ to obtain a vector point diagram (VPD), as shown in the top panel of Fig. \ref{fig3}. A CMD corresponding to each of these VPDs is plotted in the bottom panel of Figure \ref{fig3}. Proper motion is the change in the apparent positions of a celestial object when observed from the center of mass of the solar system, and each point in a VPD gives the proper motion of one star. The dense region in a VPD depicts the stars that share similar motion in space with respect to the background field stars. This is one of the primary conditions of being cluster members. Hence, we have used these stars for preliminary cluster member selection. We defined an eye estimated radius around dense region in the VPD as shown with red color in Fig. 3. Finally we calculate the the membership probability for the selected stars. The CMD in the middle of the bottom panel of Fig. \ref{fig3} appears much cleaner than the other two, reinstating the effectiveness of this primary selection method.

\subsection{Membership Probabilities }
In this section, we aim to estimate the membership probability of stars using precise measurements of PMs and parallax from Gaia-DR3. Vasilevskis et al. (1958) described the mathematical setup for determining the membership probability of stars. For NGC 5288, we have used the method devised by Balaguer-N\'{u}\~{n}ez et al. (1998) to calculate the membership probability of individual stars. This method has been used previously for both open and globular star clusters (Yadav et al. 2013, Sariya \& Yadav 2015; Sariya et al. 2021a, 2021b), and a detailed description of this method can be found in Bisht et al. (2020).

To derive the two distribution functions defined in this method, $\phi_c^{\nu}$ (cluster star distribution)
and $\phi_f^{\nu}$ (field star distribution),  we considered only those stars which have PM errors better than 1 mas~yr$^{-1}$. A group of the cluster's preliminary members shown in the VPD is found to be centered at $\mu_{xc}$=$-$3.840 mas~yr$^{-1}$, $\mu_{yc}$=$-$1.934 mas~yr$^{-1}$. We have estimated the PM dispersion for the cluster population as ($\sigma_c$) = 0.08 mas~yr$^{-1}$. For the field region, we have estimated ($\mu_{xf}$, $\mu_{yf}$) = ($-$6.18, $-$1.95) mas yr$^{-1}$ and ($\sigma_{xf}$, $\sigma_{yf}$) = (0.84, 0.70) mas yr$^{-1}$.\\

\begin{figure}[!t]
  \includegraphics[width=\columnwidth]{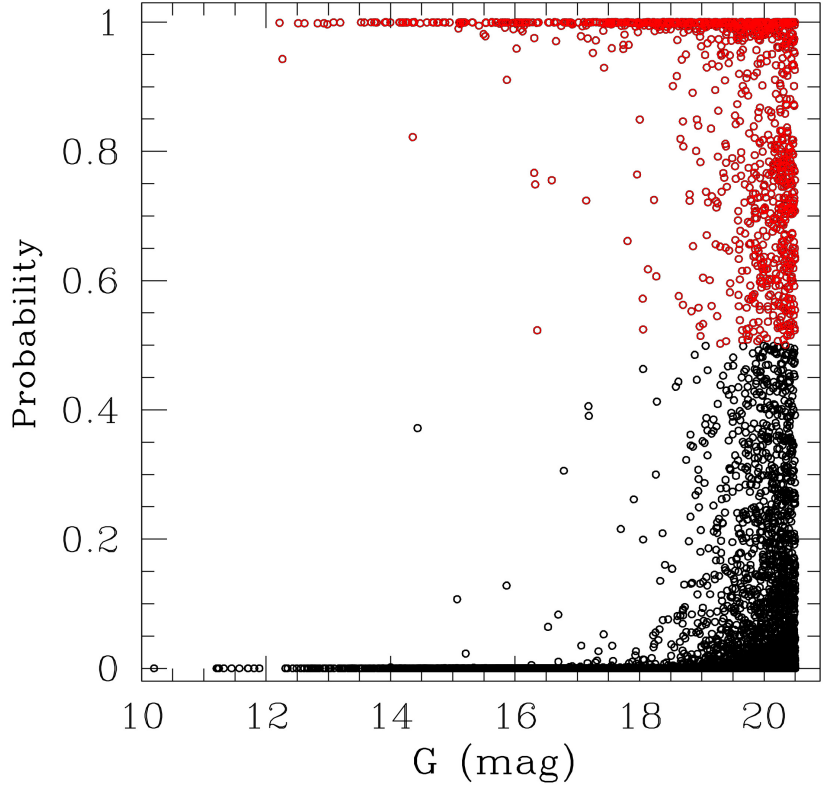}
  \caption{Plot of cluster membership probabilities against the G (mag). The red open circles represent the cluster members having membership probability greater than 50\% }
  \label{fig4}
\end{figure}

The membership probabilities ($P_{\mu}$) are thus determined, and they are shown as a function of Gaia's $G$ magnitude in Fig.~\ref{fig4}. Finally, we identified 304 stars with $P_{\mu}>$ 50\%.

To estimate the mean PM of NGC 5288, we considered the most probable cluster members and constructed the histograms for $\mu_{\alpha} cos{\delta}$ and $\mu_{\delta}$ as shown in the right panel of Fig.~\ref{fig7}. The fitting of a Gaussian function to the histograms provides the mean PM in both directions. We obtained the mean PM of NGC 5288 as $-3.840\pm0.230$ and $-1.934\pm0.162$ mas yr$^{-1}$ in $\mu_{\alpha} cos{\delta}$ and $\mu_{\delta}$ respectively. The estimated values of mean PM for this object is in very good agreement with the mean PM value given by \citet{2018A&A...618A..93C} ($\mu_{\alpha} cos{\delta}$ = $-3.850$ mas yr$^{-1}$, $\mu_{\delta}$ = $-1.935$ mas yr$^{-1}$).

\subsection{Cluster Membership Validation}
\begin{figure}[h!]
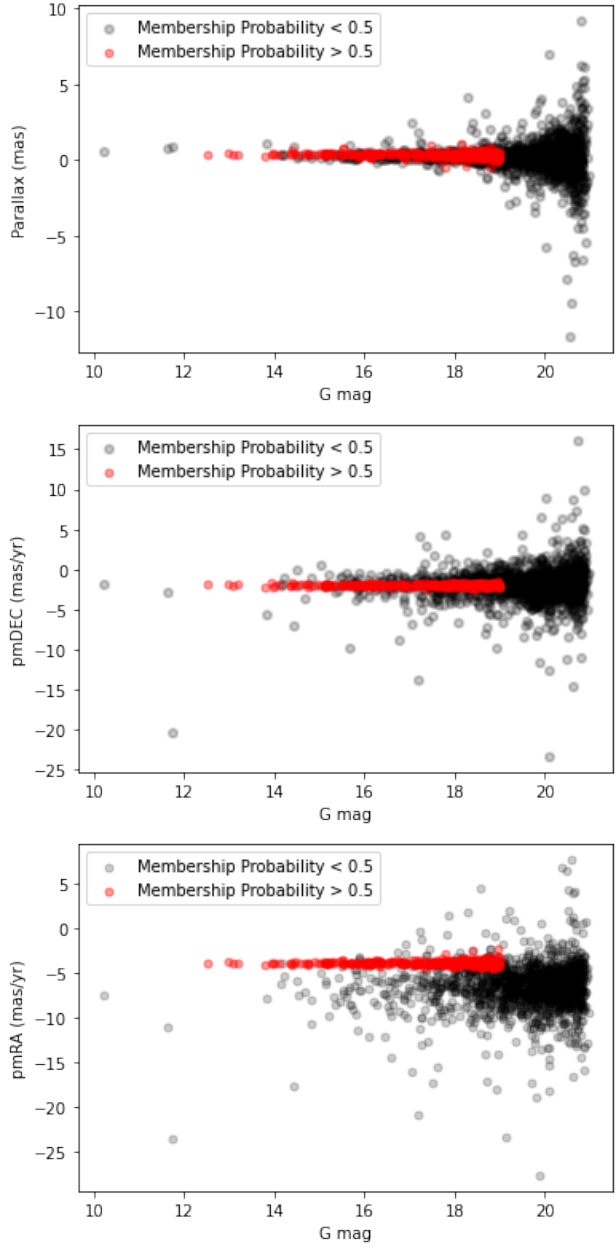

     \centering
     \begin{subfigure}[b]{1\textwidth}
         \centering
         \includegraphics[width=0.7\textwidth]{plxvsg.pdf}
     \end{subfigure}
     \begin{subfigure}[b]{1\textwidth}
         \centering
         \includegraphics[width=0.7\textwidth]{pmDEvsG.pdf}
     \end{subfigure}
     \begin{subfigure}[b]{1\textwidth}
         \centering
         \includegraphics[width=0.7\textwidth]{pmRAvsG.pdf} 
     \end{subfigure}       
    \caption{Top, middle and bottom panels represent plots of parallax distribution PM in declination and  PM in Right Ascension vs. G mag, respectively. The red dots represent the cluster members, while the black ones are the field stars.}
    \label{fig5}
\end{figure}

We have categorized the stars with membership probability greater than 50\% as cluster members, and only these stars have been used for evaluating the properties of the cluster in the current study. To validate the accuracy of our identified cluster members, we have plotted proper motions,$\mu_\alpha \cos\delta$ mas yr$^{-1}$ and $\mu_\delta$ mas yr$^{-1}$, and parallax as a function of G mag in Fig. \ref{fig5}. The graph shows a narrow distance range of parallax and proper motion coordinates for the stars categorized as cluster members, but this range is significant for field stars. Hence it is fair to say that we have successfully separated the cluster members.

\section{STRUCTURAL PROPERTIES} \label{4}
\subsection{Cluster center}
\begin{figure}[!t]
  \includegraphics[width=\columnwidth]{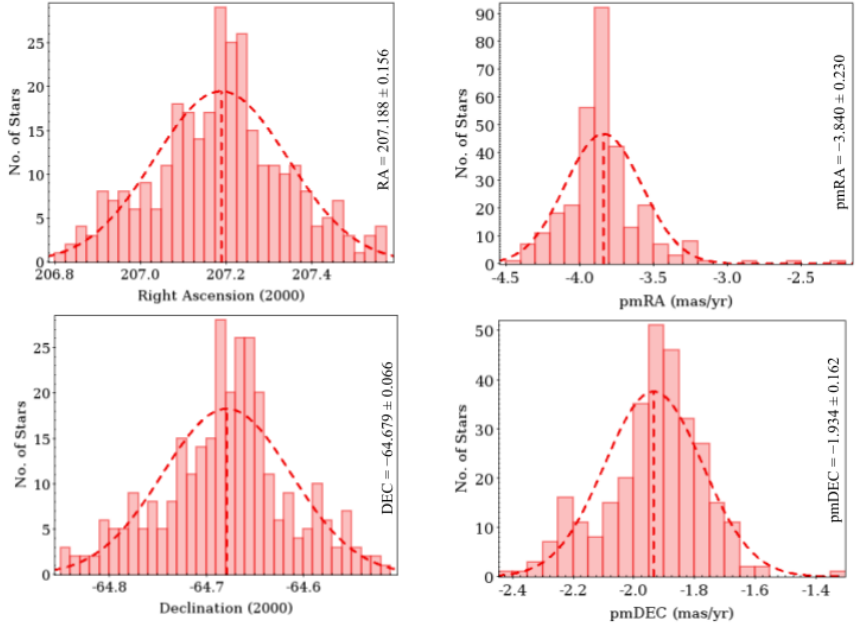}
     \caption{Histograms of star count in RA(top left), DEC(bottom left), pmRA(top right), and pmDEC(bottom left) fitted with a Gaussian. The vertical dashed-line at the Gaussian center represents the respective mean values.}
  \label{fig7}
\end{figure}

Precise estimations of the central coordinates of a cluster play a vital role in evaluating its fundamental properties. We use a star count approach that states that the cluster center is located at the point with the highest stellar density in the cluster region to determine central coordinates (Bisht et al.2021a). For this, a histogram of the star count in RA and DEC is plotted and fitted with Gaussian profiles. The mean values of RA and DEC are estimated to be $207.188$ $\pm$ $0.156$ deg ($13^h 48^m 45^s$), $-64.679$ $\pm$ $0.066$ deg ($-64^\circ 40'44''$), respectively (refer Figure \ref{fig7}). The center of the cluster is approximated to be at these coordinates. These values agree well with the values estimated in other studies, as shown in Table \ref{t3}.

\begin{figure}[!t]
  \includegraphics[width=\columnwidth]{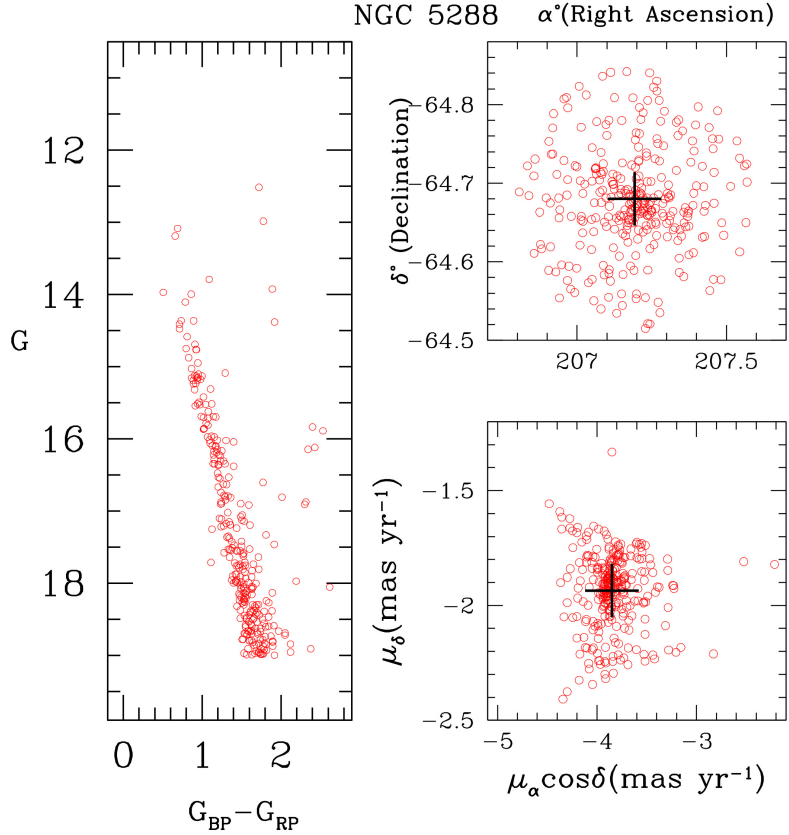}
  \caption{Plots of CMD($(G_{BP} - G_{RP}$ vs. $G$), identification chart and proper motion distribution of cluster members represented as the red dots. The plus sign denotes the cluster center.}
  \label{fig8}
\end{figure}

A CMD, identification chart and proper motion distribution of cluster members with the cluster center highlighted are plotted in Figure \ref{fig8}.
\subsection{Radial Density Profile}
\begin{figure}[!t]
  \includegraphics[width=0.9\columnwidth]{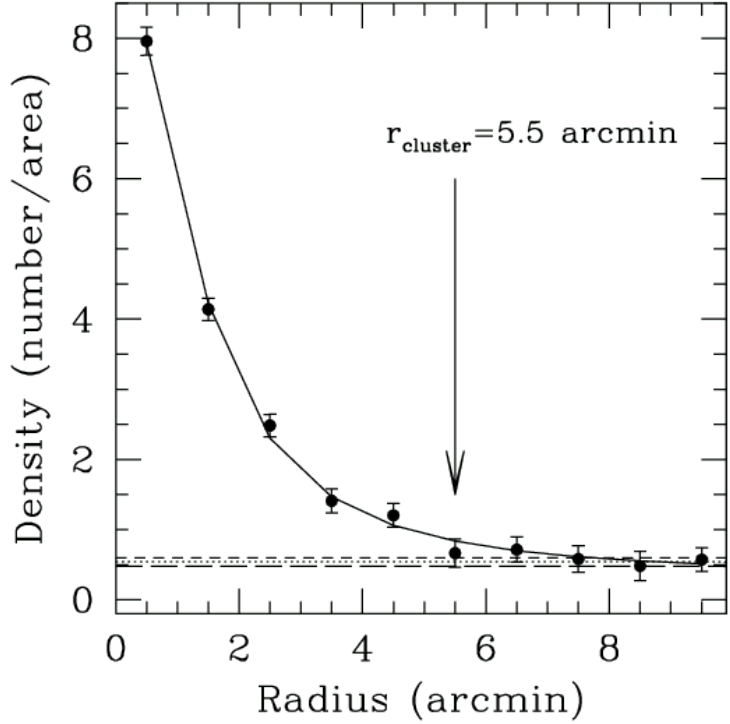}
  \caption{Radial Density Profile of NGC 5288 using cluster members. The smooth continuous line represent the fitted \citet{1962AJ.....67..471K} model. The middle dotted horizontal line depicts the background density and the long and short dashed lines represent the error background density.}
  \label{rdp}
\end{figure}

The radius of a cluster gives us a direct indication of its physical dimensions. We plotted a Radial Density Profile (RDP) of NGC 5288 as shown in Figure \ref{rdp}. We divided the cluster region into several concentric circular bins and calculated the number density, $\rho_i = \frac{N_i}{A_i}$ of each bin using cluster members. $\rho_i$ is the number of cluster members lying in the $i^{th}$ radial bin divided by the area of the $i^{th}$ bin, $A_i$. The $i^{th}$ bin corresponds to the region between $i^{th}$ and $(i-1)^{th}$ concentric circles. The observed radial profile is fitted by the \citet{1962AJ.....67..471K} model, represented as the smooth continuous line in Figure \ref{rdp}. The middle of the three horizontal lines plotted in Figure \ref{rdp} correspond to the background density, and the two lines on either side of it indicate the error in it. A mathematical expression of the model used for fitting is given as :
\begin{equation}
    f(r) = f_{bg} + \frac{f_0}{1 + \left(\frac{r}{r_c}\right)^2}
\end{equation}
where $f_{bg}, f_0, r_c$ are background density, central density and the core radius, respectively. Using the \citet{1962AJ.....67..471K} fit, we estimate the structural parameters of NGC 5288, which are given in Table \ref{t1}. We estimate the cluster radius at the point after which the cluster density merges with the field density. Following this criterion, we derived the cluster radius as $\approx 5.5$ arcmin. The previous study by \citet{2006MNRAS.367..599P} estimates the radius of NGC 5288 as $6.3$ arcmin using CCD photometry. 
We also calculate the density contrast parameter, $\delta_c$ for the cluster using the formula, $\delta_c = 1+ \frac{f_0}{f_{bg}}$ and obtain its value as $17 \pm 2$. Our estimated value of $\delta_c$ lies within the limit ($7 \le \delta_c \le 23$) given by \citet{2005A&A...431..943B}, suggesting that the cluster is compact. 
\begin{table*}[!t]\centering
  \small
  \begin{changemargin}{-1cm}{-1cm}
    \caption{Structural parameters of NGC 5288} \label{t1}
     \tablecols{5}
    \begin{tabular}{cccc c}
      \toprule
      $f_0$ (stars/arcmin$^2$) & $f_{bg}$ (stars/arcmin$^2$ )& $r_c$ (arcmin) & Cluster Radius (arcmin) &$\delta_c $ \\
      \midrule
      $28.60 \pm 0.19$& $0.54 \pm 0.06$& $1.36 \pm 0.05$ &$5.5$&$17 \pm 2$ \\ 
      \bottomrule
    \end{tabular}
  \end{changemargin}
\end{table*}

\section{Distance, Reddening and Age of NGC 5288} \label{5}
The age and distance of open clusters give us insights into the kinematics, structure, and chemical evolution of the galaxy it is situated in \citep{1993A&A...267...75F}. We have estimated distance using the parallax method and isochrone fitting method as described in this section.

\subsection{Distance using Parallax Method}
\begin{figure}[!t]
  \includegraphics[width=0.85\columnwidth]{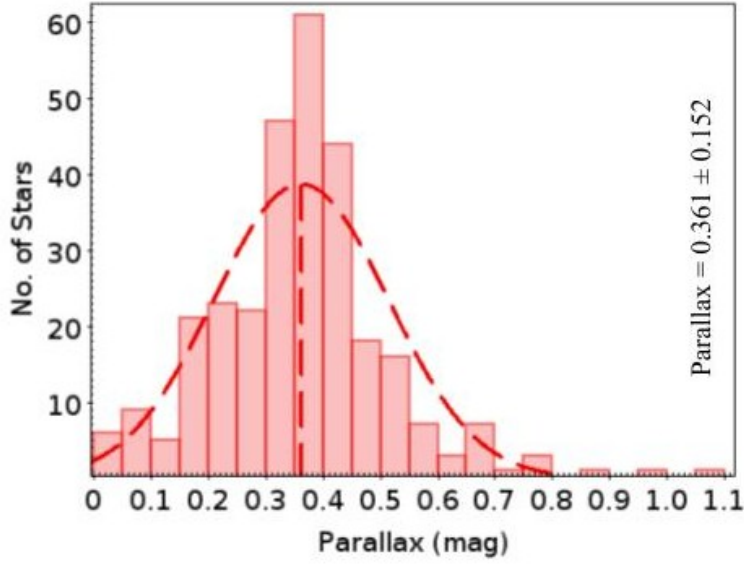}
  \caption{Histogram of parallax values of cluster members, fitted upon by a Gaussian profile with mean parallax value at $0.361$ $\pm$ $0.152$ mas indicated by a vertical line at the center of the Gaussian.}
  \label{fig9}
\end{figure}

In this section, we will use one of the essential distance measurement techniques, the parallax, to approximate the cluster's distance \citet{2018A&A...616A...9L}. A histogram of the parallax values of cluster members is plotted in which we rejected the spurious stars having negative parallax values. It is fitted to a Gaussian profile to obtain the mean parallax, which comes out to be $0.361 \pm 0.152$ mas, as shown in Figure \ref{fig9}. This value corresponds to a distance of $2.77 \pm 0.42$ kpc.  Since Gaia provides the most precise parallax values with a minimum error, this will provide the more accurate distance value. Also, the value of distances calculated from the parallax is more accurate than the values calculated from the photometric methods. Keeping this in mind, we fixed the value of cluster distance and then fitted the isochrones on the cluster main sequences to calculate the other fundamental parameters of the cluster.

\subsection{Interstellar reddening from JHK colors}
\begin{figure}[!h]
\centering
  \includegraphics[width=1\textwidth]{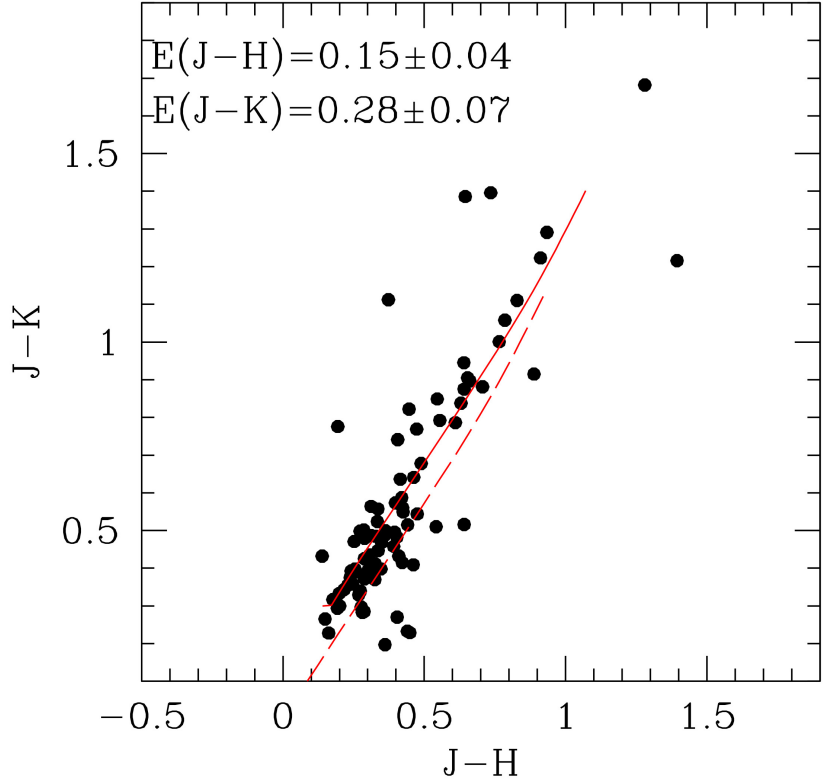}
  \caption{(J-H) vs (J-K) two color diagram. The red solid and the dashed lines are the ZAMS taken from Caldwell et al. (1993). The red dashed line is the ZAMS shifted by the value given in the text.}
  \label{cc}
\end{figure}

The cluster reddening in the near-IR region has been estimated using the (J-H) vs (J-K) color-color diagrams as shown in Figure \ref{cc}. The black dots are the cluster members, the solid red line represents the cluster's zero-age main sequence (ZAMS) taken from Caldwell et al. (1993), while the red dashed line is the ZAMS displaced by the value of E(J - H) and E(J - K), which are estimated as $0.15 \pm 0.04$ and $0.28 \pm 0.07$ respectively. The reddening is calculated using the following equations (Fiorucci \& Munari 2003) :
\begin{center}
E(J-H) = 0.309 $\times$ E(B-V) \\
E(J-K) = 0.48 $\times$ E(B-V)  
\end{center}
Using the above equations, we derived the interstellar reddening of the cluster, E(B-V) as 0.45.

\subsection{Age and Distance from Isochrone fitting}

\begin{figure}
\centering
  \includegraphics[width=1\textwidth]{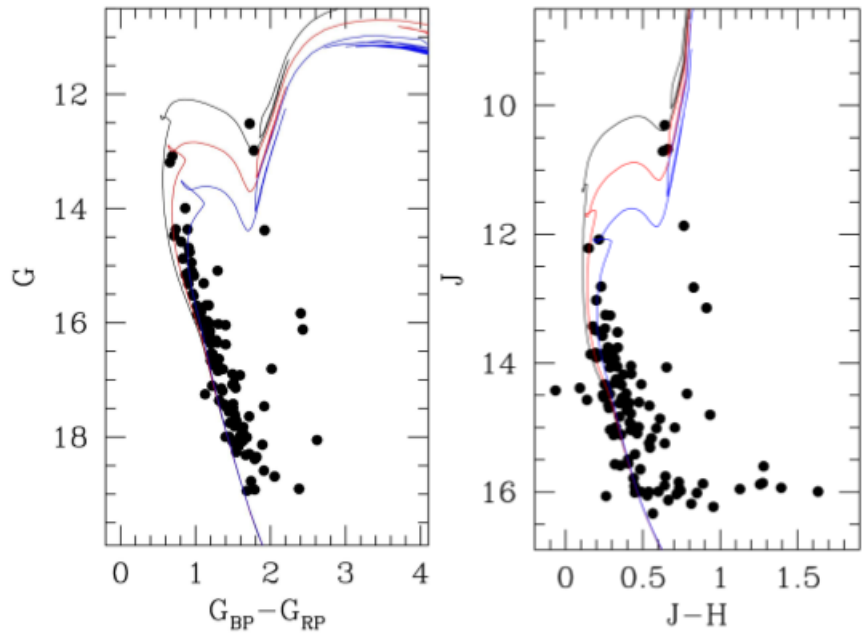}
  \caption{The (G, G$_{BP}$ - G$_{RP}$) and (J, J-H) color-magnitude diagrams of open star cluster NGC 5288. The black dots represent cluster members and the curves represent the theoretical isochrones of three different ages (8.5,8.7 and 8.9) with Z = 0.03}
  \label{fig10}
\end{figure}

We estimate important cluster parameters such as age and metallicity by fitting the theoretical isochrones of \citet{2017ApJ...835...77M} to the CMDs, (G, G$_{BP}$ - G$_{RP}$) and (J, J-H) as shown in Figure \ref{fig10}. Color-magnitude diagrams (CMDs) represent the relationship between the absolute magnitudes of the stars and their surface temperatures, usually identified by their color. CMDs are extensively used to study star clusters and are beneficial for estimating their properties (Kalirai \& Tosi 2004; Sariya et al. 2021b). We tried to fit many isochrones with different metallicity and ages and found the best fit at Z=0.03, which shows a good agreement with the literature value of 0.04 given by Piatti et al. (2006). In Figure \ref{fig10}, we have superimposed theoretical isochrones of different ages (log(age) = 8.5, 8.7 and 8.9) and Z=0.03 over the observed (G, G$_{BP}$ - G$_{RP}$) and (J, J-H) CMDs. We estimated the cluster's age to be $510 \pm 190$ Myr. Our estimated age lies in between the age given by \citet{2014yCat..35640079D} and Kharchenko et al. (2016).  We found the color excess value, $E(G_{BP}-G_{RP})$ of the cluster as 0.60. The isochrone fitting is performed so that it will result in the cluster distance close to what we estimated from parallax. From this fitting, we calculated the distance modulus (m-M) as $13.50 \pm 0.25$ mag, corresponding to a heliocentric distance of $2.64 \pm 0.11$ kpc. This value is comparable to the distance we approximated using the parallax method in the above section. The $A_G$ value for this cluster is $0.99 \pm 0.62$, which is calculated by the weighted mean method using the $A_G$ values of cluster members from the Gaia DR3 data. The value of the total to selective extinction ratio using values determined in the present analysis is compatible with the value given by Wang \& Chen (2019) within errors. A comparative analysis of the fundamental properties of cluster members derived in this study with the values in the literature is done in Table \ref{t3}.

\begin{table*}
\centering
  \small
  \begin{changemargin}{-2.5cm}{2.5cm}
    \caption{Comparison of our derived fundamental parameters of NGC 5288 with the literature values}     
    \label{t3}
     \tablecols{3}
      \setlength{\tabcolsep}{16pt}
    \begin{tabular}{ccc}   
      \toprule
      Parameters & Numerical Values & Reference \\
      \midrule
      (RA, DEC) (deg) &  $(207.188 \pm 0.156$, $-64.679 \pm 0.066)$ & Present study \\ 
         &$(207.193, -64.680)$ &  Cantat-Gaudin et al. (2018)\\
          &$(207.203, -64.673)$&  \citet{2019ApJS..245...32L}    \\
          &$(207.18333, -64.68500)$& Dias et al.(2014)\\
    $(\mu_\alpha \cos\delta$, $\mu_\delta$) (mas yr$^{-1}$)&$(-3.840 \pm 0.230, -1.934 \pm 0.162)$& Present study\\
    & $(-3.850 \pm 0.098$, $-1.935 \pm 0.087)$ & Cantat-Gaudin et al. (2018) \\
    & $(-3.841 \pm 0.169, -1.905 \pm 0.157)$&\citet{2019ApJS..245...32L} \\
    & $(-5.25 \pm 3.65, -5.79 \pm 5.58)$ & Dias et al.(2014) \\
    Cluster Radius (arcmin) & 5.5 & Present study \\
    & 2.50 & Dias et al.(2014) \\
    & 6.3 & Piatti et al.(2006) \\
    Age (Myr) & $510 \pm 190$ & present study \\
    & $1259$ &  Kharchenko et al.(2016) \\
    & $178$ & Dias et al.(2014) \\
    & $126$ & Piatti et al.(2006) \\
    Mean Parallax (mas) & $0.361 \pm 0.152$ & Present study \\
    & $0.345$ $\pm 0.041$ &  Cantat-Gaudin et al. (2018)\\
    & $0.345 \pm 0.026$ & \citet{2019ApJS..245...32L}    \\
    Distance (kpc) & $2.77 \pm 0.42$ & Present study \\
    & $2.6739$ & Cantat-Gaudin \& Anders(2019)\\
    & $5.086$ &  Kharchenko et al.(2016) \\
    & $2.158$ & Dias et al.(2014) \\
    & $2.1 \pm 0.3$ & Piatti et al.(2006) \\
    \bottomrule  
    \end{tabular}
 \end{changemargin}
\end{table*}     

\section{Luminosity and Mass Functions} \label{6}
\begin{figure}[!t]
\centering
  \includegraphics[width=0.75\textwidth]{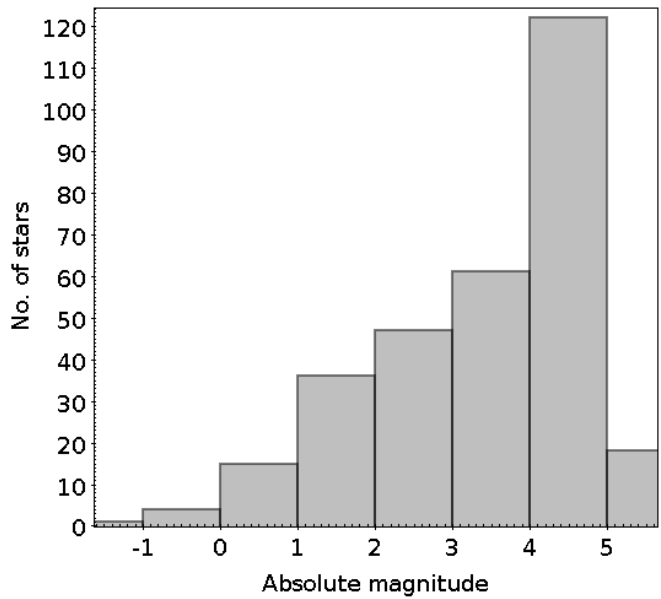}
  \caption{Luminosity function of stars in the region of NGC 5288.}
  \label{fig11}
\end{figure}

Luminosity function (LF) is the distribution of stars of a cluster in different magnitude bins. We considered cluster members with membership probability higher than 50$\%$ in $G/(G_{BP}-G_{RP})$ CMD to construct the LF for NGC 5288. To build the LF, first, we transformed the apparent $G$ magnitudes into absolute magnitudes by using the distance modulus. Then, we plotted the histogram of LF as shown in Fig \ref{fig11}. The interval of 1.0 mag was chosen to have an adequate number of stars per bin for statistical usefulness. We found an increasing LF for this cluster which is because it still retains its faint members.

\begin{figure}[!t]
\centering
\includegraphics[width=0.85\textwidth]{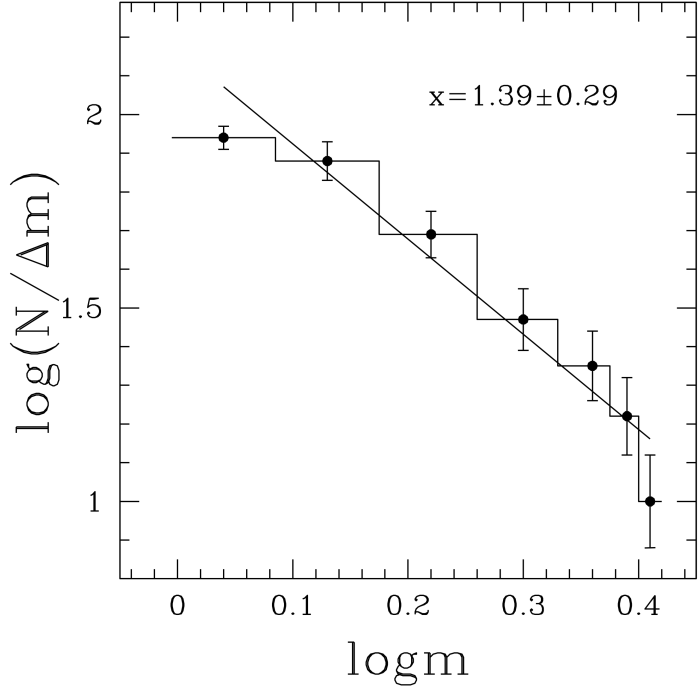}
\caption{Mass function histogram derived using the most probable members, where solid line indicates the power law given by Salpeter (1955). The error bars represent $\frac{1}{\sqrt{N}}$.}
\label{fig12}
\end{figure}

Mass function (MF) is the distribution of masses of cluster members per unit volume. A mass-luminosity connection can transform LF into the mass function (MF). Since we could not acquire an observational transformation, we must rely on theoretical models. To convert LF into MF, we used cluster parameters derived in this paper, and theoretical models are given by Marigo et al. (2017). The resulting MF is shown in Fig.~\ref{fig12}. The mass function slope can be derived from the linear relation\\

\begin{equation}
{\rm log}\frac{dN}{dM} = -(1+x) \times {\rm log}(M)+constant\\
\end{equation}
In the above relation, $dN$ symbolizes the number of stars in a mass bin $dM$ with the central mass $M$, and $x$ is the mass function slope. The Salpeter (1955) value for the mass function slope is $x=1.35$. This form of Salpeter demonstrates that the number of stars in each mass range declines rapidly with increasing mass. Our derived MF slope value, $x=1.39\pm0.29$, agrees with Salpeter's slope within uncertainty. Utilizing this value of mass function slope within the mass ranges 1.0~-~2.7 $M_{\odot}$, the total and mean mass for the cluster members were acquired as $\sim 464 M_{\odot} and \sim1.53 M_{\odot}$, respectively.

\subsection{Dynamical relaxation time}
In the lifetime of star clusters, encounters between its member stars gradually lead to increased energy equipartition throughout the clusters. In this process, massive stars are concentrated towards the cluster core and transfer their kinetic energy to fainter stars. The relaxation time $T_{R}$ is defined as the time in which the stellar velocity distribution becomes Maxwellian and expressed by the following formula:\\

\begin{equation}
~~~~~~~~~~~T_{R}=\frac{8.9\times10^5\sqrt{N}\times{R_{h}}^{3/2}}{\sqrt{\bar{m}}\times log(0.4N)}\\
\end{equation}

where $N$ represents the number of stars in the clusters (in our case, the ones with membership probability higher than 50$\%$), $R_{h}$ is the cluster half mass radius expressed in parsec and $\bar{m}$ is the average mass of the cluster members (Spitzer \& Hart 1971) in the solar unit. The value of $R_{h}$ is assumed as 2.56 pc, which is equal to half of the cluster's extent. Using the above formula, the value of dynamical relaxation time $T_{R}$ is determined as $24.6$ Myr, significantly less than the cluster's age. Hence, we conclude that NGC 5288 is a dynamically relaxed cluster.


\section{Orbit of the Cluster} \label{7}
Galactic orbits are valuable tools to explore the dynamical aspects of the objects in our Galaxy. We derived orbits and orbital parameters of NGC 5288 using Galactic potential models discussed by
Irrgang et al. (2013). Galactic potentials used in the present analysis are explained by Rangwal et al. (2019). In this analysis, we adopted the refined parameters of Galactic potentials from Bajkova \& Bobylev (2017); they used newly available observational data to update these parameters. Information required for orbital integration such as center coordinates ($\alpha$ and $\delta$), mean proper motions ($\mu_{\alpha}cos\delta$, $\mu_{\delta}$), parallax angles, age and heliocentric distance ($d_{\odot}$) have been taken from the present analysis. Value for the radial velocity of the cluster, which is equal to
$-23.83\pm1.49$ km/sec is taken from Soubiran et al. (2018).

\begin{table*}[!t]\centering
  \small
  \begin{changemargin}{1.4cm}{1.4cm}
    \caption{} \label{position}
     \tablecols{9}
    \begin{tabular}{cccccc}
      \toprule
      R (kpc)& Z(kpc) & $\phi$ & U (km/s) & V (km/s) & W (km/s)  \\
      \midrule
      6.919 & -0.101 & 0.320 & 30.490 & -243.025 & 5.457 \\
      \bottomrule
    \end{tabular}
    \small \item{\textit{Table 3: Position and velocity components of NGC 5288 in Galactocentric coordinate system. The meaning of the symbols used are described in the text}}
  \end{changemargin}
\end{table*}

We adopted the right-handed coordinate system to convert equatorial position and velocity components into Galactic position (($R$, $\phi$, $z$) and Galactic-space velocity components ( $U, V, W$), where ($R$, $U$), ($\phi$, $V$), and ($z$, $W$) are radial, tangential, and vertical space and velocity components, respectively. Here, the x-axis is taken positive towards the Galactic center, the y-axis is towards the direction of Galactic rotation, and the z-axis is in the vicinity of the Galactic north pole. The value of
Galactic center and North-Galactic pole is adopted from Reid \& Brunthaler (2004) as ($17^{h}45^{m}32^{s}.224, -28^{\circ}56^{\prime}10^{\prime\prime}$) and ($12^{h}51^{m}26^{s}.282, 27^{\circ}7^{\prime}42^{\prime\prime}.01$) respectively. To implicate a correction for Standard Solar Motion and motion of the Local Standard of Rest (LSR), we utilized the Sun's position and space-velocity components as ($8.3,0,0.02$) kpc and ($11.1, 12.24, 7.25$) km/s (Schonrich et al. 2010) respectively.

\begin{table*}[!t]\centering
\small
 \begin{changemargin}{-1.4cm}{-1.4cm}
    \caption{}\label{orbitpara}
     \tablecols{9}
    \begin{tabular}{cccccccc}
      \toprule
     $ e$ & $R_{a}$ & $R_{p}$ & $Z_{max}$ & $E$ $(100km/sec)^{2})$ & $L_{z}$ $(100km/sec)^{2})$ & $T_{R}$ (Myr) & $T_{Z}$ (Myr) \\
      \midrule
       -0.010 & 7.213 & 7.363 & 0.125 & -12.146 & -16.816 &  177.621 & 68.837 \\
      \bottomrule
        \end{tabular}
    \small \item{\textit{Table 4: Orbital properties of NGC 5288. Here e is eccentricity, $R_{p}$ is the maximum distance travelled by the cluster in the x direction, $R_{a}$ is the maximum distance travelled by the cluster in the y direction, $E$ is energy, $L_{z}$ is momentum, $T_{R}$ is time period for motion in plane, and $T_{Z}$ is time period for motion in the z-direction.}} 
   
  \end{changemargin}
\end{table*}

\begin{figure*}
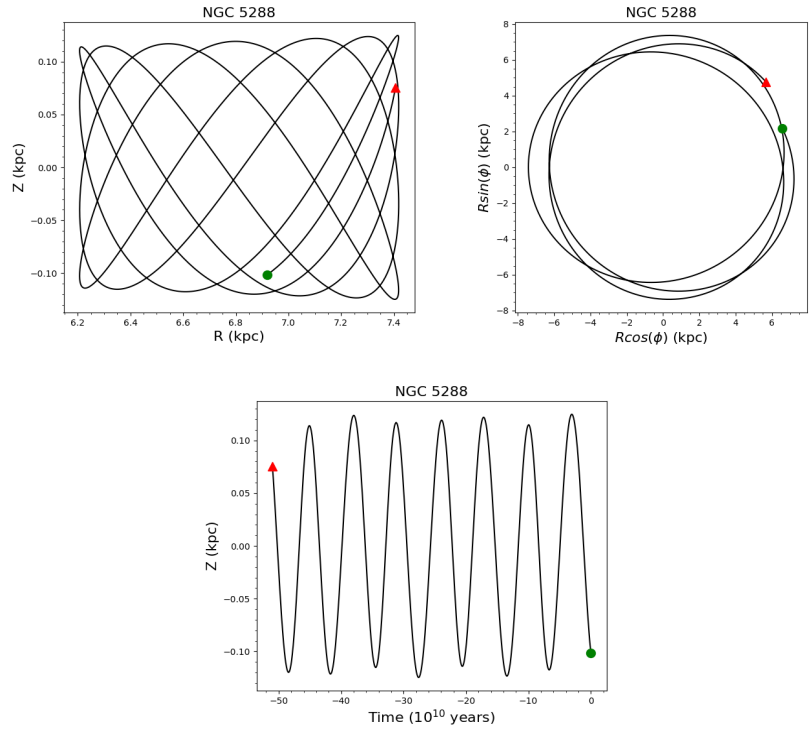

\centering
  \includegraphics[width=6 cm, height=5 cm]{n5288_1.pdf}
  \includegraphics[width=5 cm, height=5 cm]{n5288_2.pdf}
  \includegraphics[width=6 cm, height=5 cm]{n5288_3.pdf}
  \caption{The Galactic orbits of NGC 5288 were obtained with the Galactic potential model described by Rangwal et al. (2019). The time of integration is taken equal to the age of the cluster. The top left panel shows the side view, and the top middle panel shows the top view of the orbits. With time, the bottom panel shows the cluster's motion in the Galactic disk. The filled red triangles and the green circles denote the birth and the present-day positions of NGC 5288 in the Galaxy.}
  \label{orbit_fig}
\end{figure*}

The resultant orbits are shown in Fig. \ref{orbit_fig}. In the top left panel of the figure, we plotted the path followed by the cluster in $R$ and $Z$ plane, representing the orbit's side view. The top right panel is a plot between $x$ and $y$ components of the Galactic distance $R$. The bottom panel shows a plot between the time of orbit integration and the object's height from the Galactic disc. The red-filled triangle and green-filled circle denote the cluster's birth and present-day position, respectively. The orbital parameters are summarised in Table \ref{orbitpara}. 

From Fig \ref{orbit_fig} and Table \ref{orbitpara}, it is evident that the cluster is following a boxy pattern and moving in a circular orbit around the Galactic center. The cluster is a part of Galactic thin disc, moving inside the solar circle (the position of the Sun is at 8.3 kpc). The cluster was born very close to the Galactic disc and hence highly affected by the tidal forces originating from the disc, which led to a small-scale orbit height. This also led to a smaller time period in the vertical direction, as visible from Table \ref{orbitpara} and the bottom panel of figure \ref{orbit_fig}. The cluster is very young, so it must contain its fainter stars, but we can notice a significant dip in the luminosity function of the cluster (Fig. \ref{fig11}), which reflects an absence of very low mass stars in the cluster. We expect this because of the close proximity of the cluster with the Galactic disc. This cluster is highly affected by the Galactic tidal force and has lost its low mass stars to the field.

\section{Conclusion} \label{8}
We have investigated a poorly studied Open Cluster, NGC 5288, using 2MASS and Gaia DR3 photometric and astrometric database. We estimated the membership probabilities of stars using their PM and parallax measurements and identified 304 cluster members with membership probability greater than 50\%. We estimated the fundamental properties of the cluster, investigated its structure, conducted a dynamical study and derived its galactic orbit and orbital parameters. A summary of the major outcomes from this study are as follows-
\begin{enumerate}
    \item The cluster center is estimated as $\alpha = $207.188 $\pm$ 0.156 deg ($13^h 48^m 45^s$) and $\delta =$ -64.679 $\pm$ 0.066 deg ($-64^\circ 40'44''$) using the cluster members. 
    \item The core radius and the cluster radius are estimated as $1.36 \pm 0.05$ arcmins and 5.5 arcmins, respectively, using the radial density profile. 
    \item The mean proper motions of cluster members in Right Ascension and Declination are calculated as $(-3.840 \pm 0.230)$ and $(-1.934 \pm 0.162)$ mas yr$^{-1}$ respectively. The interstellar reddening, E(B-V) of the cluster is estimated as 0.45
    \item The heliocentric distance of the cluster is estimated using the parallax method and is equal to $2.77 \pm 0.42$ kpc. We determine the cluster's age as $510 \pm 190$ Myr by fitting the obtained CMDs of the cluster with theoretical isochrones with Z=0.03, given by Marigo et al.(2017). 
    \item The mass function slope of $1.39\pm0.29$  is found for NGC 5288 to be in fair agreement within uncertainty with the value (1.35) given by Salpeter (1955). Our study also indicates that this cluster is dynamically relaxed.
    \item We integrated the orbits of the cluster for a time duration equal to the cluster's age. We found that the cluster is born near the Galactic 
    disc and moves in a circular orbit with a very small scale height.
    We also expect that the absence of low-mass stars of the cluster is an effect of strong tidal forces originating from the Galactic disc.
    
\end{enumerate}

\section{Acknowledgements}
The authors are thankful to the anonymous referee for useful comments, which improved the contents of the paper significantly.
This work has made use of data from the European Space Agency (ESA) mission {\it Gaia} (\url{https://www.cosmos.esa.int/gaia}), processed by the {\it Gaia} Data Processing and Analysis Consortium (DPAC, \url{https://www.cosmos.esa.int/web/gaia/dpac/consortium}). Funding for the DPAC has been provided by national institutions, in particular, the institutions participating in the {\it Gaia} Multilateral Agreement. This work has used TOPCAT \url{http://www.starlink.ac.uk/topcat}. It was developed mostly in the UK within various UK and Euro-VO projects (Starlink, AstroGrid, VOTech, AIDA, GAVO, GENIUS, DPAC) and under PPARC and STFC grants. Its underlying table processing facilities were provided by the related packages STIL and STILTS. This research has made use of the VizieR catalogue access tool, CDS, Strasbourg, France (DOI : 10.26093/cds/vizier). The original description of the VizieR service was published in 2000, A\&AS 143, 23". 
This work has also made use of SAOImageDS9, an image visualization tool for astronomical data. This research has also made use of SIMBAD database operated at CDS, Strasbourg, France.

\end{document}